\documentclass[aps,prl,preprint,groupedaddress]{revtex4-1}

\usepackage{graphicx}				
\usepackage[utf8]{inputenc}				
\usepackage{axodraw4j}
\usepackage{xspace}
\usepackage{amssymb}
\def\eg{\emph{e.g.}}

\newcommand{\herwig}{H\protect\scalebox{0.8}{ERWIG}7\xspace}
\newcommand{\pythia}{P\protect\scalebox{0.8}{YTHIA}8\xspace}
\def\thau{\tilde{\tau}}
\def\Paragraph#1{\paragraph{#1 ---\hspace{-2mm}}}

\def\showcommentsflag{0}
\newcommand{\showcomments}{\def\showcommentsflag{1}}

\newcounter{commentcounter}%
\newcommand{\comment}[1]{\ifnum\showcommentsflag > 0%
\addtocounter{commentcounter}{1}%
{{\Red{\ensuremath{\ddagger^{\arabic{commentcounter}}}}}}%
\marginpar{\raggedright\tiny\it{{\Red{\ensuremath{\ddagger^{\arabic{commentcounter}}}}} {#1}}}
\fi%
}
\newcommand{\commentdel}[2]{\ifnum\showcommentsflag > 0%
\Red{\sout{#1}}\comment{#2}%
\fi
}
\newcommand{\commentadd}[2]{\ifnum\showcommentsflag > 0%
\comment{#2}\Red{#1}%
\else
#1
\fi
}
\newcommand{\commentchange}[3]{\ifnum\showcommentsflag > 0%
\Red{\sout{#2}}\comment{#3}\Red{#1}%
\else
#1
\fi
}
\newcommand{\nocomment}[1]{\ifnum\showcommentsflag > 0%
{\tiny\it\Red{\{#1}\}}
\fi%
}
\newcommand{\nocommentdel}[1]{\ifnum\showcommentsflag > 0%
\Red{\sout{#1}}%
\fi
}
\newcommand{\nocommentadd}[1]{\ifnum\showcommentsflag > 0%
\Red{#1}%
\else
#1
\fi
}
\newcommand{\nocommentchange}[2]{\ifnum\showcommentsflag > 0%
\Red{\sout{#2}}\Red{#1}%
\else
#1
\fi
}

\showcomments

\begin{document}
\title{Collectivity without plasma in hadronic collisions}
\author{Christian Bierlich}
\author{G{\"o}sta Gustafson}
\author{Leif L{\"o}nnblad}
\affiliation{Lund University}
\thanks{This work was funded in part by the Swedish Research Council, contracts number 2016-03291, 2016-05996 and 2017-0034, in part by the European Research Council (ERC) under the European Union’s Horizon 2020 research and innovation programme, grant agreement No 668679, and in part by the MCnetITN3 H2020 Marie Curie Initial Training Network, contract 722104.}
\preprint{LU-TP 17-33}
\preprint{MCnet-17-16}

\date{\today}

\begin{abstract}

  We present a microscopic model for collective effects in high
  multiplicity proton--proton collisions, where multiple partonic
  subcollisions give rise to a dense system of strings. From lattice
  calculations we know that QCD strings are transversely extended, and
  we argue that this should result in a transverse pressure and
  expansion, similar to the flow in a deconfined plasma. The model is
  implemented in the \pythia Monte Carlo event generator, and we find
  that it can qualitatively reproduce the long range azimuthal
  correlations forming a near-side ridge in high multiplicity
  proton--proton events at LHC energies.
  
\end{abstract}

\maketitle
\Paragraph{Introduction}%
The general features of proton-proton collisions, such as jets,
multiplicity distributions, and (approximate) particle ratios, can be
described by dynamical models based on string
\cite{Andersson:1983ia} or cluster
\cite{Marchesini:1983bm} hadronisation, \eg\ \pythia
\cite{Sjostrand:2006za,Sjostrand:2014zea} and \herwig
\cite{Bellm:2015jjp}, in a very satisfactory way. In contrast, heavy
ion collisions show collective features interpreted as flow in a
deconfined, thermalized plasma \cite{Ollitrault:1992bk,Ackermann:2000tr}. 
Nucleus collisions also show higher rates for strangeness 
\cite{Baechler:1991pp}. These features have been interpreted as 
indicating fundamentally different dynamics in the two systems.

There are, however, also many similarities between collisions in small
and in large systems. Many features in nucleus collisions, such as
multiplicity distributions and particle distributions in rapidity and
$p_\perp$, could fairly well be described by early models based on
non-interacting strings (\eg\ DPM \cite{Capella:1983ie} and Fritiof
\cite{Andersson:1986gw}). On the other hand recent precise
measurements at the LHC show flow-like effects also in pp and p$A$
collisions
\cite{Aaboud:2016yar,Khachatryan:2016txc,Abelev:2012ola}. They also
show increasing strangeness and baryon rates in pp events with high
multiplicity \cite{ALICE:2017jyt}.  This has raised the question if a
QGP is formed also in small collision systems. Conversely, one could
instead ask if collective effects in nucleus collisions could possibly
originate from non-thermal interactions between string-like colour
fields. This would entail a picture where collective phenomena does
not arise from the formation of a deconfined QGP state, but rather as
an emergent behaviour of dense configurations of confined QCD flux
tubes.  A third possibility is if the two pictures coexist, with a
dense thermalised central ``core'' and an outer ``corona''. Such a
picture is implemented in the quite successful EPOS model
\cite{Werner:2007bf,Pierog:2013ria}.

It was early suggested that the many strings in an $AA$ collision may
interact coherently as ``ropes''~\cite{Biro:1984cf}. Rope formation or
``percolation'' have subsequently been studied by several authors, see
\eg\ refs.~\cite{Bialas:1984ye,Gyulassy:1986jq,Merino:1991nq,Sorge:1992ej,
  Amelin:1994mc,Braun:2001us,Soff:2002bn,Capella:2006fw,
  Braun:2012kn,Braun:2014ica,Bierlich:2014xba}. Generally these analyses have
predicted higher rates for strangeness and baryons, and larger transverse
momenta, due to the stronger field in the rope.

The high energy density in overlapping strings also ought to give a transverse pressure,
resulting in a transverse expansion seen as a transverse flow. This
should give not only enhanced transverse momenta, in particular for high mass
particles, but would also give rise to angular correlations. Such
correlations were early considered by Abramovsky \emph{et
  al.}~\cite{Abramovsky:1988zh}, and a Monte Carlo ``toy model''
studying this effect in PbPb collisions was presented in
\cite{Altsybeev:2015vma}. 

A high string density can also be reached in pp and p$A$ collisions at
high enough energies, allowing for rope formation also in these
smaller systems.
In ref.~\cite{Bierlich:2016vgw} we presented a proof of principle for
a flow-like transverse expansion, due to overlapping strings in high
energy pp collisions. In this letter we want use the concept of
overlapping strings to study this effect more thoroughly, and compare
the results with experimental data. To that effect we have implemented
the resulting model in the \pythia event generator, and find good
agreement between the model and the pp "ridge" first observed by the
CMS collaboration \cite{Khachatryan:2010gv}.

\Paragraph{The Lund string}%
The confining force field between coloured partons is in the Lund
string model \cite{Andersson:1983ia} described by a ``massless
relativistic string'', which represents an idealised picture of a flux tube
with no transverse extension. Like a
linear electric field, the string is invariant under longitudinal
boosts, and has no momentum in the direction of the string (apart from
the force on the endpoints). Gluons are treated as point-like
transverse excitations on the string \cite{Andersson:1979ij}, and for
a set of colour connected partons, moving apart from a common origin,
the string is stretched from a quark via the colour-ordered gluons to
an antiquark.

\begin{figure}
\scalebox{0.5}{\mbox{\begin{picture}(451,226) (0,32)
    \SetWidth{1.0}
    \Bezier[dash,dsize=0.969](248.177,86.28)(279.199,87.25)(302.466,102.761)(333.488,133.783)
    \Bezier[dash,dsize=0.969](248.177,86.28)(217.155,87.25)(193.889,102.761)(155.111,133.783)
    \Bezier[dash,dsize=0.969](155.111,133.783)(124.089,164.805)(100.822,188.072)(62.044,226.85)
    \Bezier[dash,dsize=0.969](333.488,133.783)(387.777,180.316)(426.555,219.094)(426.555,219.094)
    \Line[arrow,arrowpos=0.5,arrowlength=5,arrowwidth=2,arrowinset=0.2,flip](100.822,180.316)(248.177,32.961)
    \Line[arrow,arrowpos=0.5,arrowlength=5,arrowwidth=2,arrowinset=0.2](248.177,32.961)(395.533,180.316)
    \Line[arrow,arrowpos=0.5,arrowlength=5,arrowwidth=2,arrowinset=0.2](100.822,180.316)(104.7,184.679)
    \Line[arrow,arrowpos=0.5,arrowlength=5,arrowwidth=2,arrowinset=0.2](395.533,180.316)(390.201,186.133)
    \Line[arrow,arrowpos=0.5,arrowlength=5,arrowwidth=2,arrowinset=0.2,flip](104.7,184.679)(147.355,142.993)
    \Line[arrow,arrowpos=0.5,arrowlength=5,arrowwidth=2,arrowinset=0.2](147.355,142.993)(152.687,147.84)
    \Line[arrow,arrowpos=0.5,arrowlength=5,arrowwidth=2,arrowinset=0.2,flip](152.687,147.84)(195.343,104.7)
    \Line[arrow,arrowpos=0.5,arrowlength=5,arrowwidth=2,arrowinset=0.2](195.343,104.7)(206.976,116.818)
    \Line[arrow,arrowpos=0.5,arrowlength=5,arrowwidth=2,arrowinset=0.2,flip](206.976,116.818)(238.968,83.857)
    \Line[arrow,arrowpos=0.5,arrowlength=5,arrowwidth=2,arrowinset=0.2](238.968,83.857)(263.688,109.062)
    \Line[arrow,arrowpos=0.5,arrowlength=5,arrowwidth=2,arrowinset=0.2,flip](263.688,109.062)(279.199,92.582)
    \Line[arrow,arrowpos=0.5,arrowlength=5,arrowwidth=2,arrowinset=0.2](279.199,92.582)(321.855,133.783)
    \Line[arrow,arrowpos=0.5,arrowlength=5,arrowwidth=2,arrowinset=0.2,flip](321.855,133.783)(329.611,126.028)
    \Line[arrow,arrowpos=0.5,arrowlength=5,arrowwidth=2,arrowinset=0.2](329.611,126.028)(390.201,186.133)
    \Line[arrow,arrowpos=1,arrowlength=5,arrowwidth=2,arrowinset=0.2](104.7,184.679)(72.708,224.911)
    \Line[arrow,arrowpos=1,arrowlength=5,arrowwidth=2,arrowinset=0.2](152.687,147.84)(116.333,195.827)
    \Line[arrow,arrowpos=1,arrowlength=5,arrowwidth=2,arrowinset=0.2](206.976,116.818)(186.133,164.805)
    \Line[arrow,arrowpos=1,arrowlength=5,arrowwidth=2,arrowinset=0.2](263.688,109.062)(271.444,149.294)
    \Line[arrow,arrowpos=1,arrowlength=5,arrowwidth=2,arrowinset=0.2](321.855,133.783)(356.755,180.316)
    \Line[arrow,arrowpos=1,arrowlength=5,arrowwidth=2,arrowinset=0.2](390.201,186.133)(449.821,257.872)
    \LongArrow(60,40)(80,40)
    \LongArrow(60,40)(60,60)
    \Text(53,60)[c]{\LARGE $t$}
    \Text(80,30)[c]{\LARGE $x$}
    \Text(250,60)[c]{\Large $A/\kappa^2$}
  \end{picture}}}

  \caption{\label{fig:fan}Breakup of a string between a quark and an
    antiquark in a $x-t$ diagram. New $q\bar{q}$ pairs are produced
    around a hyperbola, and combine to the outgoing hadrons. The
    original $q$ and $\bar{q}$ move along light-like trajectories. The area
    enclosed by the quark lines is the coherence area $A$ in
    eq.~(\ref{eq:arealaw}), in units of the string tension $\kappa$.
  }
\end{figure}
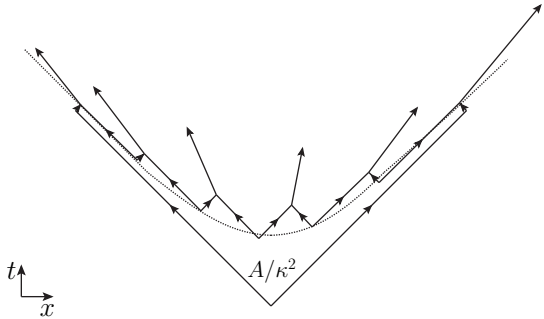

The most simple situation is a quark and an antiquark produced in an
$e^+e^-$ annihilation, and moving apart. A string is stretched from
the common production point, and breaks into pieces, hadrons, by
repeated production of $q\bar{q}$ pairs, as sketched in
fig.~\ref{fig:fan}. The boost invariant string is reflected in a
boost invariant distribution of the hadrons. The spacetime separations
between the production points are space-like, and in each frame the
hadrons with the smallest energies in this specific frame, are
produced first in time. 

The probability for a specific final state is given by \cite{Andersson:1983jt}
\begin{equation}
d\mathcal{P}\propto \exp(-bA)\times d\Omega.
\label{eq:arealaw}
\end{equation}
Here $bA$ is (the imaginary part of) the action for the relativistic
string, with $A$ equal to the space-time area (in units of the string
tension $\kappa$) covered by the string before its breakup into
hadrons (see fig.~\ref{fig:fan}). For a single hadron species with
mass $m$, the $n$ particle phase space $\Omega$ (in 1+1 dimensions) is
given by $d\Omega=\prod_{i=1}^n
[N\,d^2p_i\,\delta(p_i^2-m^2)]\,\delta^{(2)}(\sum p_i -
P_{\mathrm{tot}})$. The parameter $N$ here specifies the relation
between the phase space for $n+1$ and $n$ particles.

For a straight string the area $A$ is easily expressed in terms of the hadron
momenta, and the result can be generated by 
successively peeling off the hadrons from the quark or antiquark
ends. Here each hadron is given a fraction $z$ of the remaining
light-cone momentum, determined by the distribution (in case of a
single hadron species with mass $m$)
\begin{equation}
    f(z)=N z^a \exp(- b\, m^2/z).
\label{eq:splitting}
\end{equation}
The three parameters $N$, $a$, and $b$ are related by normalisation, leaving
two parameters to be determined by experiments.

The breakup points for the string are located around a hyperbola in
spacetime, with a typical proper time given by
\begin{equation}
\langle \tau^2 \rangle= \frac{1+a}{b\, \kappa^2},
\label{eq:Gamma}
\end{equation}
where $\kappa$ is the tension of the string. With values $a=0.68$,
$b=0.98$~GeV$^{-2}$ (the default values in \pythia), and
$\kappa=0.9-1$~GeV/fm, we obtain the typical breaking time around
2~fm. Thus the string breaks typically when the original quark and
antiquark are about 4~fm apart. The string then breaks in two
pieces, which move apart keeping their size, with the new antiquark
trailing after the initial quark increasing its energy due to the pull
from the string, as illustrated in fig.~\ref{fig:fan}. Eventually the
string breaks again, and in the successive breaks the string pieces
become smaller and smaller.

\Paragraph{Interactions between strings}%
We note that just after the production of the initial $q\bar{q}$ pair,
the colour field is necessarily compressed, not only longitudinally
but also transversely. Also in a high energy pp collision the strings
are stretched between charges emerging from a very limited region
within two Lorentz contracted pancakes. As illustrated in
fig.~\ref{fig:expansion}, the flux tube expands both longitudinally and
transversely with the speed of light. Two neighbouring flux tubes will
then start to overlap and interact close to $z=0$ (where $z$ is the
longitudinal coordinate) in the specific frame used in the analysis,
as this is where the flux tube expands most rapidly. As seen in
fig.~\ref{fig:fan} this is also the region where those particles are
produced, which are slow in this particular frame.

\newsavebox{\firstbox}
\savebox{\firstbox}(100,60)[bl]{
  \GOval(45,30)(1,1)(0){0}
  \GOval(55,30)(1,1)(0){0}
  \LongArrow(45,30)(35,30)
  \LongArrow(55,30)(65,30)
  \Line(45,30)(50,35)
  \Line(45,30)(50,25)
  \Line(55,30)(50,35)
  \Line(55,30)(50,25)
  \Line(48,30)(50,32)
  \Line(48,30)(50,28)
  \Line(52,30)(50,32)
  \Line(52,30)(50,28)
  \Line(49.5,30)(50,30.5)
  \Line(50.5,30)(50,30.5)
  \Line(49.5,30)(50,29.5)
  \Line(50.5,30)(50,29.5)}

\newsavebox{\secondbox}
\savebox{\secondbox}(100,60)[bl]{
  \GOval(40,30)(1,1)(0){0}
  \GOval(60,30)(1,1)(0){0}
  \LongArrow(40,30)(30,30)
  \LongArrow(60,30)(70,30)
  \Line(40,30)(50,40)
  \Line(40,30)(50,20)
  \Line(60,30)(50,40)
  \Line(60,30)(50,20)
  \Line(43,30)(50,37)
  \Line(43,30)(50,23)
  \Line(57,30)(50,37)
  \Line(57,30)(50,23)
  \Line(46,30)(50,34)
  \Line(46,30)(50,26)
  \Line(54,30)(50,34)
  \Line(54,30)(50,26)
  \Line(49,30)(50,31)
  \Line(51,30)(50,31)
  \Line(49,30)(50,29)
  \Line(51,30)(50,29)}

\newsavebox{\thirdbox}
\savebox{\thirdbox}(100,60)[bl]{
  \GOval(35,30)(1,1)(0){0}
  \GOval(65,30)(1,1)(0){0}
  \LongArrow(35,30)(25,30)
  \LongArrow(65,30)(75,30)
  \Line(35,30)(45,40)
  \Line(35,30)(45,20)
  \Line(65,30)(55,40)
  \Line(65,30)(55,20)
  \Line(45,40)(55,40)
  \Line(45,20)(55,20)
  \Line(38,30)(45,37)
  \Line(38,30)(45,23)
  \Line(62,30)(55,37)
  \Line(62,30)(55,23)
  \Line(45,37)(55,37)
  \Line(45,23)(55,23)
  \Line(41,30)(45,34)
  \Line(41,30)(45,26)
  \Line(59,30)(55,34)
  \Line(59,30)(55,26)
  \Line(45,34)(55,34)
  \Line(45,26)(55,26)
  \Line(44,30)(45,31)
  \Line(44,30)(45,29)
  \Line(56,30)(55,31)
  \Line(56,30)(55,29)
  \Line(45,31)(55,31)
  \Line(45,29)(55,29)}
\newsavebox{\fourthbox}
\savebox{\fourthbox}(100,60)[bl]{
  \GOval(30,30)(1,1)(0){0}
  \GOval(70,30)(1,1)(0){0}
  \LongArrow(30,30)(20,30)
  \LongArrow(70,30)(80,30)
  \Line(30,30)(40,40)
  \Line(30,30)(40,20)
  \Line(70,30)(60,40)
  \Line(70,30)(60,20)
  \Line(40,40)(60,40)
  \Line(40,20)(60,20)
  \Line(33,30)(40,37)
  \Line(33,30)(40,23)
  \Line(67,30)(60,37)
  \Line(67,30)(60,23)
  \Line(40,37)(60,37)
  \Line(40,23)(60,23)
  \Line(36,30)(40,34)
  \Line(36,30)(40,26)
  \Line(64,30)(60,34)
  \Line(64,30)(60,26)
  \Line(40,34)(60,34)
  \Line(40,26)(60,26)
  \Line(39,30)(40,31)
  \Line(39,30)(40,29)
  \Line(61,30)(60,31)
  \Line(61,30)(60,29)
  \Line(40,31)(60,31)
  \Line(40,29)(60,29)}
\newsavebox{\firstboxr}
\savebox{\firstboxr}(100,60)[bl]{
  \GOval(45,30)(1,1)(0){0}
  \GOval(55,30)(1,1)(0){0}
  \LongArrow(45,30)(35,30)
  \LongArrow(55,30)(65,30)
  \SetColor{Red}
  \Line(45,30)(50,35)
  \Line(45,30)(50,25)
  \Line(55,30)(50,35)
  \Line(55,30)(50,25)
  \Line(48,30)(50,32)
  \Line(48,30)(50,28)
  \Line(52,30)(50,32)
  \Line(52,30)(50,28)
  \Line(49.5,30)(50,30.5)
  \Line(50.5,30)(50,30.5)
  \Line(49.5,30)(50,29.5)
  \Line(50.5,30)(50,29.5)}

\newsavebox{\secondboxr}
\savebox{\secondboxr}(100,60)[bl]{
  \GOval(40,30)(1,1)(0){0}
  \GOval(60,30)(1,1)(0){0}
  \LongArrow(40,30)(30,30)
  \LongArrow(60,30)(70,30)
  \SetColor{Red}
  \Line(40,30)(50,40)
  \Line(40,30)(50,20)
  \Line(60,30)(50,40)
  \Line(60,30)(50,20)
  \Line(43,30)(50,37)
  \Line(43,30)(50,23)
  \Line(57,30)(50,37)
  \Line(57,30)(50,23)
  \Line(46,30)(50,34)
  \Line(46,30)(50,26)
  \Line(54,30)(50,34)
  \Line(54,30)(50,26)
  \Line(49,30)(50,31)
  \Line(51,30)(50,31)
  \Line(49,30)(50,29)
  \Line(51,30)(50,29)}

\newsavebox{\thirdboxr}
\savebox{\thirdboxr}(100,60)[bl]{
  \GOval(35,30)(1,1)(0){0}
  \GOval(65,30)(1,1)(0){0}
  \LongArrow(35,30)(25,30)
  \LongArrow(65,30)(75,30)
  \SetColor{Red}
  \Line(35,30)(45,40)
  \Line(35,30)(45,20)
  \Line(65,30)(55,40)
  \Line(65,30)(55,20)
  \Line(45,40)(55,40)
  \Line(45,20)(55,20)
  \Line(38,30)(45,37)
  \Line(38,30)(45,23)
  \Line(62,30)(55,37)
  \Line(62,30)(55,23)
  \Line(45,37)(55,37)
  \Line(45,23)(55,23)
  \Line(41,30)(45,34)
  \Line(41,30)(45,26)
  \Line(59,30)(55,34)
  \Line(59,30)(55,26)
  \Line(45,34)(55,34)
  \Line(45,26)(55,26)
  \Line(44,30)(45,31)
  \Line(44,30)(45,29)
  \Line(56,30)(55,31)
  \Line(56,30)(55,29)
  \Line(45,31)(55,31)
  \Line(45,29)(55,29)}
\newsavebox{\fourthboxr}
\savebox{\fourthboxr}(100,60)[bl]{
  \GOval(30,30)(1,1)(0){0}
  \GOval(70,30)(1,1)(0){0}
  \LongArrow(30,30)(20,30)
  \LongArrow(70,30)(80,30)
  \SetColor{Red}
  \Line(30,30)(40,40)
  \Line(30,30)(40,20)
  \Line(70,30)(60,40)
  \Line(70,30)(60,20)
  \Line(40,40)(60,40)
  \Line(40,20)(60,20)
  \Line(33,30)(40,37)
  \Line(33,30)(40,23)
  \Line(67,30)(60,37)
  \Line(67,30)(60,23)
  \Line(40,37)(60,37)
  \Line(40,23)(60,23)
  \Line(36,30)(40,34)
  \Line(36,30)(40,26)
  \Line(64,30)(60,34)
  \Line(64,30)(60,26)
  \Line(40,34)(60,34)
  \Line(40,26)(60,26)
  \Line(39,30)(40,31)
  \Line(39,30)(40,29)
  \Line(61,30)(60,31)
  \Line(61,30)(60,29)
  \Line(40,31)(60,31)
  \Line(40,29)(60,29)}
\newsavebox{\firstboxb}
\savebox{\firstboxb}(100,60)[bl]{
  \GOval(45,30)(1,1)(0){0}
  \GOval(55,30)(1,1)(0){0}
  \LongArrow(45,30)(35,30)
  \LongArrow(55,30)(65,30)
  \SetColor{Blue}
  \Line(45,30)(50,35)
  \Line(45,30)(50,25)
  \Line(55,30)(50,35)
  \Line(55,30)(50,25)
  \Line(48,30)(50,32)
  \Line(48,30)(50,28)
  \Line(52,30)(50,32)
  \Line(52,30)(50,28)
  \Line(49.5,30)(50,30.5)
  \Line(50.5,30)(50,30.5)
  \Line(49.5,30)(50,29.5)
  \Line(50.5,30)(50,29.5)}

\newsavebox{\secondboxb}
\savebox{\secondboxb}(100,60)[bl]{
  \GOval(40,30)(1,1)(0){0}
  \GOval(60,30)(1,1)(0){0}
  \LongArrow(40,30)(30,30)
  \LongArrow(60,30)(70,30)
  \SetColor{Blue}
  \Line(40,30)(50,40)
  \Line(40,30)(50,20)
  \Line(60,30)(50,40)
  \Line(60,30)(50,20)
  \Line(43,30)(50,37)
  \Line(43,30)(50,23)
  \Line(57,30)(50,37)
  \Line(57,30)(50,23)
  \Line(46,30)(50,34)
  \Line(46,30)(50,26)
  \Line(54,30)(50,34)
  \Line(54,30)(50,26)
  \Line(49,30)(50,31)
  \Line(51,30)(50,31)
  \Line(49,30)(50,29)
  \Line(51,30)(50,29)}

\newsavebox{\thirdboxb}
\savebox{\thirdboxb}(100,60)[bl]{
  \GOval(35,30)(1,1)(0){0}
  \GOval(65,30)(1,1)(0){0}
  \LongArrow(35,30)(25,30)
  \LongArrow(65,30)(75,30)
  \SetColor{Blue}
  \Line(35,30)(45,40)
  \Line(35,30)(45,20)
  \Line(65,30)(55,40)
  \Line(65,30)(55,20)
  \Line(45,40)(55,40)
  \Line(45,20)(55,20)
  \Line(38,30)(45,37)
  \Line(38,30)(45,23)
  \Line(62,30)(55,37)
  \Line(62,30)(55,23)
  \Line(45,37)(55,37)
  \Line(45,23)(55,23)
  \Line(41,30)(45,34)
  \Line(41,30)(45,26)
  \Line(59,30)(55,34)
  \Line(59,30)(55,26)
  \Line(45,34)(55,34)
  \Line(45,26)(55,26)
  \Line(44,30)(45,31)
  \Line(44,30)(45,29)
  \Line(56,30)(55,31)
  \Line(56,30)(55,29)
  \Line(45,31)(55,31)
  \Line(45,29)(55,29)}
\newsavebox{\fourthboxb}
\savebox{\fourthboxb}(100,60)[bl]{
  \GOval(30,30)(1,1)(0){0}
  \GOval(70,30)(1,1)(0){0}
  \LongArrow(30,30)(20,30)
  \LongArrow(70,30)(80,30)
  \SetColor{Blue}
  \Line(30,30)(40,40)
  \Line(30,30)(40,20)
  \Line(70,30)(60,40)
  \Line(70,30)(60,20)
  \Line(40,40)(60,40)
  \Line(40,20)(60,20)
  \Line(33,30)(40,37)
  \Line(33,30)(40,23)
  \Line(67,30)(60,37)
  \Line(67,30)(60,23)
  \Line(40,37)(60,37)
  \Line(40,23)(60,23)
  \Line(36,30)(40,34)
  \Line(36,30)(40,26)
  \Line(64,30)(60,34)
  \Line(64,30)(60,26)
  \Line(40,34)(60,34)
  \Line(40,26)(60,26)
  \Line(39,30)(40,31)
  \Line(39,30)(40,29)
  \Line(61,30)(60,31)
  \Line(61,30)(60,29)
  \Line(40,31)(60,31)
  \Line(40,29)(60,29)}

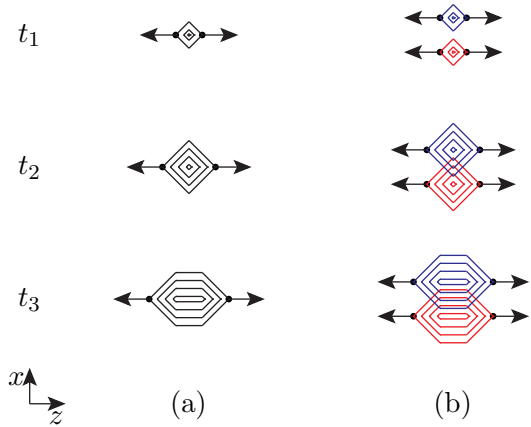
\begin{figure}
  \begin{picture}(200,150)(0,150)
    \put(20,260){\usebox{\firstbox}}
    \put(120,253.5){\usebox{\firstboxr}}
    \put(120,266.5){\usebox{\firstboxb}}
    \Text(10,290)[c]{$t_1$}

    \put(20,210){\usebox{\secondbox}}
    \put(120,203.5){\usebox{\secondboxr}}
    \put(120,216.5){\usebox{\secondboxb}}
    \Text(10,240)[c]{$t_2$}
    \put(20,160){\usebox{\thirdbox}}
    \put(120,153.5){\usebox{\thirdboxr}}
    \put(120,166.5){\usebox{\thirdboxb}}
    \Text(10,190)[c]{$t_3$}

    \LongArrow(10,150)(20,150)
    \LongArrow(10,150)(10,160)
    \Text(5,160)[c]{$x$}
    \Text(20,145)[c]{$z$}
    \Text(70,150)[c]{(a)}
    \Text(170,150)[c]{(b)}

  \end{picture}
  \caption{\label{fig:expansion}(a) The forcefield between a quark and
    an antiquark separating from a common origin expands both
    longitudinally and transversely, until the transverse extension
    saturates.  (b) The expansion is boost invariant. Therefore, in any
    frame two parallel flux tubes begin to overlap and interact in the
    centre in the specific frame chosen.}
\end{figure}

The repulsion gives the flux tubes a transverse velocity, a process
which we will refer to as \textit{shoving}. In a pp collision the
density of strings is not too high, and the time until breakup
($\tau\sim2$~fm) is large compared to both the width of the flux tubes
and the radius of the proton (both $<1$~fm). We therefore expect
that the force field has (almost) reached its equilibrium transverse
width, before it breaks up into hadrons. These equilibrium flux tubes
may be a single triplet string, or a rope stretched between higher
colour multiplets, as discussed in ref.~\cite{Bierlich:2014xba}.

The different time scales imply that the process has \emph{three
  separate phases}: i) An initial phase where the individual flux tubes
expand but still do not interact. ii) A second phase where the
flux tubes interact, repel each other, and reach equilibrium triplet
strings or ropes specified by definite SU(3) multiplets. iii) A final
phase in which the flux tubes break into hadrons. Naturally the
boundaries between the phases are not sharp, but we will in this paper
assume that they can be treated separately.

\Paragraph{Flux tube repulsion}%
Colour flux tubes are similar to vortex lines in a superconductor. In
a type I superconductor there is a homogenous field within the tube
(similar to the confined field in the bag model), while in a type II
superconductor the field falls off exponentially outside a thin
core. Lattice calculations indicate that the properties of a
colour flux tube lie in between type I and II superconductors, with
field strength close to a Gaussian \cite{Cea:2014uja}.

If the energy in the flux tube would be dominated by a longitudinal
colour-electric field, the interaction between them would be given by their
overlap. For Gaussian distributions this would give a force per unit string
length 
\begin{equation}
\frac{\delta F(d_\perp)}{\delta z} \equiv f(d_\perp) = \frac{g\kappa
  d_\perp}{R^2}\exp\left(-\frac{d^2_\perp(t)}{4R^2} \right).
\label{eq:repulsion}
\end{equation}
Here $d_\perp$ is the transverse separation between the flux tubes,
$R$ is the (time dependent) width of the energy distribution in a
single flux tube, and $\kappa$ the tension in a single string. 
In eq.~(\ref{eq:repulsion}) we have included a tunable parameter $g$, and the
situation with a dominating colour-electric field would correspond to $g=1$. 
We note, however, that in the bag model (similar to a type I
superconductor) half of the string energy corresponds to destroying
the condensate inside the flux tube, and in a type II superconductor
the larger part of the energy lies in the current keeping the flux
together. Due to these uncertainties we expect that $g$ might deviate from 1,
possibly within an order of
magnitude. We motivated the Gaussian form in eq.~(\ref{eq:repulsion})
by lattice results, but the end result is more sensitive to the value of $g$, 
than the shape of the field.

For a boost invariant system it is convenient to introduce hyperbolic
coordinates
\begin{equation}
\thau = \sqrt{t^2 - z^2}, \,\,\,\,\, \eta = \ln ((t+z)/\thau).
\label{eq:hyper}
\end{equation}
(Note that $\thau$ is boost invariant, but not equal to the eigentime, given
by $\tau=\sqrt{t^2-z^2-\mathbf{x}_\perp^2}$.)  Near $z=0$ we get $\delta z=t\, 
\delta \eta$, and the force in eq.~(\ref{eq:repulsion}) gives $dp_\perp/dt\,
\delta z=f(d_\perp)$. (For non-relativistic velocities this gives the
acceleration $d v_\perp/dt = \delta z \, f(d_\perp)/(\delta z\,\kappa)
=f(d_\perp)/\kappa$.)  Boost invariance then gives the two equations
\begin{equation}
\frac{d p_\perp}{\thau d\thau\, d\eta} =
f(d_\perp),\,\,\,\,\frac{d^2
  d_\perp}{d\thau^2} = \frac{f(d_\perp)}{\kappa}. 
\label{eq:dpt}
\end{equation}
When the flux tubes have separated, the boost invariant total $p_\perp$ per
unit $\eta$ is given by integration of eqs.~(\ref{eq:dpt}).

\Paragraph{Hadronisation}%
The dynamics of the relativistic string is well described in
ref.~\cite{Artru:1979ye}.
The motion of a string stretched such that $d\,p_\perp/d\,\eta$ is 
constant and along the $x$ direction, is described by a smoothly bent curve
in space-time (for $t\ge0$)
\begin{equation}
x_\mu = \Bigl(\thau \cosh(\eta);\, \alpha\, \text{arcsh}(\thau/\alpha),\, 0,\,
\thau \sinh(\eta)\Bigr). 
\label{eq:stringstate}
\end{equation}
The transverse momentum is here specified by the parameter $\alpha =  
(d p_\perp/d\eta) /\kappa$.
We note that the transverse coordinate $x$ is independent of $\eta$,
and the result is explicitly boost invariant. The expression in
eq.~(\ref{eq:stringstate}) does not explicitly
depend on the total energy W, which only shows up in the
kinematical boundaries $0<|\sigma|< W/2$ and $t< W/2$, after which time the
string ends are pulled back again. 

We note that the string state discussed here is very different from a
straight string boosted transversely with some velocity $v_\perp$. The
directions of the initial quark and antiquark would then be rotated an
angle $\theta=\text{arcsine}\, v_\perp$, and the produced hadrons
would be limited to the rapidity range between plus and minus
$\log (\cot(\theta/2))$. The distribution would thus not be boost
invariant, but depend on the Lorentz frame used for the transverse
boost.

When calculating the breakup into hadrons, of a string described in
eq.~(\ref{eq:stringstate}), we realise that when the
string is not straight, the relation between the space-time area $A$ in
eq.~(\ref{eq:arealaw}) and the final hadron momenta in the splitting function
in eq.~({\ref{eq:splitting}) will be modified. We plan to discuss how to solve
this problem in a coming paper (inspired by earlier work in ref. 
\cite{Andersson:1998sw}}), and we will in the implementation used in
this letter apply an approximate method, described in the
next paragraph. 

\Paragraph{Monte Carlo Implementation}%
We have used the \pythia event generator \cite{Sjostrand:2014zea} to
study the effect of string repulsion, and we simulate the shoving mechanism by
adding soft gluons to the strings. As the string interaction depends on their
positions in impact parameter space, and since the models for multi-parton
interaction (MPI) and parton showers in \pythia make no attempt to
calculate parton positions in transverse space, we have devised and
implemented a simple model for this. Here we assume a 2D Gaussian mass
distribution of the protons, 
and pick the transverse coordinates for each MPI according to the
convolution of the mass distributions at the impact parameter of the
collision\footnote{The shoving model has been implemented in
\pythia, which can be obtained at \texttt{http://home.thep.lu.se/Pythia}. 
The implementation allows for the user to supply their own model of MPI 
distribution in impact parameter space.}.

In a real pp collision many strings are stretched between partons
forming minijets, or occasional hard jets. The string pieces between
these partons, which we call (colour) \emph{dipoles}, will not be fully
aligned with the beam axis. In our current implementation the system
is separated in segments along the beam axis, which are presently 0.1
rapidity units wide. One string may here overlap with several other
strings, and within each segment the shoving is implemented as the sum
of many small kicks between all pairs of overlapping string
pieces. This is done in several time steps. After each time
step, the string configurations are updated, and the process continues
until hadronisation time is reached ($\tau\sim$2~fm after the initial
collision). We allow the partons to move a finite
time ($\sim$1~fm) before they begin to interact with each other. 
This also accounts for the fact that it takes some time for the strings to
extend transversely, before they feel the transverse repulsion.

For technical reasons \pythia cannot handle very soft gluons
properly. Thus, if the state contains a group of colour-connected
gluons with invariant mass below a cut ($\sim$0.2~GeV), they will in
the program be replaced by a single, harder, gluon.
The result of this substitution is that the extra transverse momentum
\textit{per unit rapidity}, due to the shoving, is correctly
reproduced. However, the hadron multiplicity will be artificially
somewhat increased, which implies that the effect on $p_\perp$
\textit{per particle} will be underestimated. Such an increase in
multiplicity would also give an artificial widening of a jet and to
avoid this we also abstain from adding excitations to string segments
which already before the shoving has a high $p_\perp>2$~GeV.

\begin{figure}
  \includegraphics[width=0.5\textwidth]{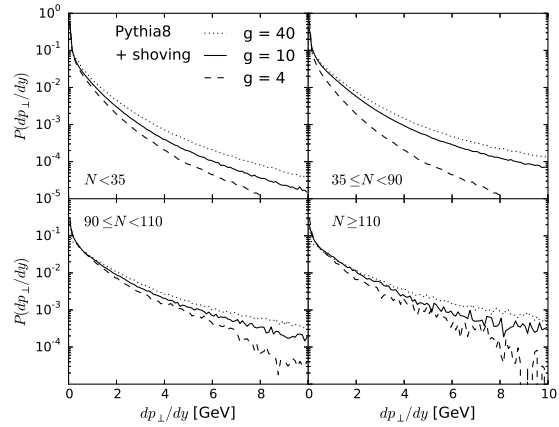}
  \caption{\label{fig:modelplot} The extra transverse momentum per unit
    rapidity, $dp_\perp/dy$ (measured in the interval $|y|<2.5$), for 
    four different final state particle multiplicities, and three different
    values for the shoving parameter $g$.}
\end{figure}

In fig. \ref{fig:modelplot} we show the extra transverse momentum
per unit rapidity due to shoving for particles within the region $|y|<2.5$
(matching 
the experimental acceptance in ref. \cite{Khachatryan:2010gv}). The
result is presented in four final-state particle multiplicity bins,
and for three values of the parameter $g$ determining the strength of
the repulsion between strings in eq.~(\ref{eq:repulsion}). As discussed
there, we expect that a reasonable value for $g$ would be 1 within an order
of magnitude. In fig. \ref{fig:modelplot} we show results for $g=4$, and
$g=10$. For illustration we also show results for $g=40$, which we
regard as unreasonably large. We note that as expected the extra
$dp_\perp/dy$ increases both with the strength parameter $g$
and with multiplicity. We also note that even for the lowest value,
$g=4$, the added $p_\perp$ per unit rapidity can reach quite large
values.

\Paragraph{Flow-like effects}%
We mentioned above that the problem with hadronising soft gluons implies that
the multiplicity is somewhat overestimated in events with low multiplicity. We
do, however, not expect a noticeable effect from this problem for two-particle
correlations. This expectation has been checked by comparing two-particle
correlations for default \pythia and for shoving with $g = 0.01$. We here note
that although the  multiplicity increases when adding even very small
excitations, these excitations do not affect the two-particle
correlations. Thus there is no appearance of a ridge for this small $g$-value.
In fig. \ref{fig:modelplot} we also see that the extra $p_\perp$ grows as
expected, with increasing event multiplicity. 

\begin{figure}
  \includegraphics[width=0.5\textwidth]{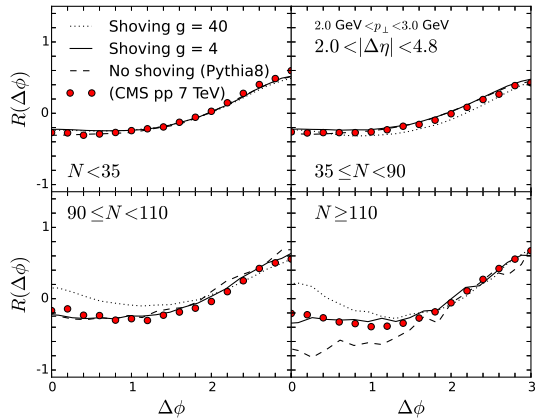}
  \caption{\label{fig:correlations} 
	   Di-hadron correlation functions for pp collisions at 7~TeV, in four
           centrality intervals, for 
	   two values of the shoving parameter $g$, compared to default \pythia.
	   For $g=4$, adding shoving produces a ridge similar to the data from
           CMS \cite{Khachatryan:2010gv}.}
\end{figure}

To calculate the ``ridge effect'' we employ an analysis similar to the
one chosen by experiments \cite{Khachatryan:2010gv}, where a signal
distribution $S(\Delta\phi,\Delta\eta)$ is divided by a random
background distribution, $B(\Delta\phi,\Delta\eta)$, constructed by
combining particles from two different events in the same centrality
class. In fig.~\ref{fig:correlations} we show results for correlations in the
range $2< |\Delta \eta|<4.8$ for two
values of the shoving parameter $g$, and compare with default \pythia,
where no ridge is expected.  Data from CMS \cite{Khachatryan:2010gv}
is added to the figure, but the comparison was carried out by
ourselves, and is not guaranteed to include all experimental
corrections. It would be beneficial for further model development to
have the analysis available in Rivet~\cite{Buckley:2010ar}. We note
that the ridge appearing at high event multiplicity is qualitatively
reproduced with a value for the interaction strength $g=4$.

\Paragraph{Conclusion}%
We have shown, that long-range two-particle correlations in pp
collisions can be accounted for in a model, where the strings interact
as overlapping flux tubes. The assumed interaction potential is inspired by 
lattice calculations.
The model is implemented in \pythia and contains a
single free parameter, which determines the strength of the
interaction. We show that for a physically reasonable value of this
parameter, the model qualitatively reproduces the "ridge" structure in
high multiplicity pp collisions, as observed by several experiments.

We wish to emphasise that the model does not rely on the assumption of
a deconfined plasma, normally employed by thermodynamical models of
collective effects. On the contrary the collective behaviour is a
consequence of the confining fields, resulting in an interaction
between the strings that is without diffusion or loss of energy. Thus,
although the string system is not deconfined nor thermalised, the
transverse expansion has important similarities with the expansion of
a boost-invariant perfect (non-viscous) liquid. 

In a coming publication we want to improve the approximations in the
implementation of the "shoving model" presented here, and combine it with 
the rope hadronisation model in ref.~\cite{Bierlich:2014xba}.
Our plan is then to include these effects in our model for collisions
with nuclei~\cite{Bierlich:2016smv}, to see if they can adequately
describe data showing collective effects in these larger
systems. Would such a comparison turn out successful, this would
challenge the current paradigm in heavy ion physics. It would then be
necessary to find observables sensitive to dynamical differences
between the traditional approach assuming a thermalised plasma, and
the non-thermalised dynamics described here.

\bibliography{shoving}

\end{document}